%% file: 4U1812-12_Period.tex
\documentclass[twocolumn]{aastex631}
\newcommand{\Four}{4U\,1812--12}
\newcommand{\Porb}{$P_{\rm orb}$}
\usepackage{soul}

\include{definitions}
\shorttitle{A tentative 114-minute orbital period for \Four}
\shortauthors{Armas Padilla et al.}
\graphicspath{{./}{figures/}}

\begin{document}

\title{A tentative 114-minute orbital period challenges the ultra-compact nature of the X-ray binary \Four}

\author[0000-0002-4344-7334]{Montserrat Armas Padilla}
\affiliation{Instituto de Astrofísica de Canarias, E-38205 La Laguna, Tenerife, Spain}
\affiliation{Departamento de Astrofísica, Universidad de La Laguna, E-38206 La Laguna, Tenerife, Spain}
\email{m.armaspadilla@iac.es}

\author[0000-0002-4717-5102]{Pablo Rodríguez-Gil}
\affiliation{Instituto de Astrofísica de Canarias, E-38205 La Laguna, Tenerife, Spain}
\affiliation{Departamento de Astrofísica, Universidad de La Laguna, E-38206 La Laguna, Tenerife, Spain}

\author[0000-0002-3348-4035]{Teo Muñoz-Darias}
\affiliation{Instituto de Astrofísica de Canarias, E-38205 La Laguna, Tenerife, Spain}
\affiliation{Departamento de Astrofísica, Universidad de La Laguna, E-38206 La Laguna, Tenerife, Spain}

\author[0000-0002-5297-2683]{Manuel A.P.Torres}
\affiliation{Instituto de Astrofísica de Canarias, E-38205 La Laguna, Tenerife, Spain}
\affiliation{Departamento de Astrofísica, Universidad de La Laguna, E-38206 La Laguna, Tenerife, Spain}

\author[0000-0001-5031-0128]{Jorge Casares}
\affiliation{Instituto de Astrofísica de Canarias, E-38205 La Laguna, Tenerife, Spain}
\affiliation{Departamento de Astrofísica, Universidad de La Laguna, E-38206 La Laguna, Tenerife, Spain}

\author[0000-0002-0092-3548]{Nathalie Degenaar}
\affiliation{Anton Pannekoek Institute for Astronomy, University of Amsterdam, Science Park 904, 1098 XH Amsterdam, The Netherlands}

\author[0000-0003-4236-9642]{Vik S. Dhillon}
\affiliation{Department of Physics and Astronomy, University of Sheffield, Sheffield S3 7RH, UK}
\affiliation{Instituto de Astrofísica de Canarias, E-38205 La Laguna, Tenerife, Spain}

\author[0000-0003-3944-6109]{Craig~O. Heinke}
\affiliation{Department of Physics and Astronomy, University of Alberta, Edmonton T6G 2E9, Alberta, Canada}

\author[0000-0001-7221-855X]{Stuart P. Littlefair}
\affiliation{Department of Physics and Astronomy, University of Sheffield, Sheffield S3 7RH, UK}

\author[0000-0002-2498-7589]{Thomas R. Marsh}
\affiliation{Department of Physics, University of Warwick, Coventry CV4 7AL, UK}



\begin{abstract}

We present a detailed time-resolved photometric study of the ultra-compact X-ray binary candidate \Four. The multicolor light curves obtained with HiPERCAM on the 10.4-m Gran Telescopio Canarias show a $\simeq 114$~min modulation similar to a superhump. Under this interpretation, this period should lie very close to the orbital period of the system. Contrary to what its other observational properties suggest (namely, persistent dim luminosity, low  optical-to-X-ray flux ratio and lack of hydrogen features in the optical spectrum), this implies that \Four\ is most likely not an ultra-compact X-ray binary, which are usually defined as systems with orbital periods lower than 80~min. We discuss the nature of the system, showing that a scenario in which \Four\ is the progenitor of an ultra-compact X-ray binary may reconcile all the observables.

\end{abstract}



\section{Introduction} \label{sec:intro}

The ultra-compact family of low mass X-ray binaries (LMXBs) is composed of those systems in which a compact object, a neutron star or a black hole, accretes material from an evolved, hydrogen-deficient companion star in a tight orbit with a period \Po~$<80$~min \citep{Rappaport1982,Verbunt1995}. Three evolutionary channels, depending on the nature of the donor star, have been proposed to explain the origin of such compact systems: the white dwarf channel, the helium star channel and the evolved main-sequence star channel \citep[see e.g.,][]{Nelemans2010b, Heinke2013}. Distinguishing between these formation paths is not always straightforward, since the final state of the donor is similar for the three scenarios. In this regard, the study of the progenitors of ultra-compact systems (i.e., before the orbital period becomes shorter than 80~min) may be key to shed light on the origin of this family.

The neutron star LMXB \Four\ is a strong ultra-compact candidate. Its persistently low X-ray luminosity ($\sim4\times10^{35}$~$\lum$) and low optical--to--X-ray flux ratio suggest that it harbors a small accretion disk \citep{Paradijs1994, Bassa2006, IntZand2007}. In addition, its optical spectrum lacks hydrogen spectral features, which suggest a hydrogen-exhausted donor star \citep{ArmasPadilla2020}.

In order to confirm the ultra-compact nature of \Four, we performed a detailed time-resolved photometric study using data taken with the HiPERCAM imager on the 10.4-m Gran Telescopio Canarias (GTC).

\section{Observations and reduction}\label{sec:Obs}

We obtained images of \Four\ with 
HiPERCAM \citep{dhillonetal21-1} on the GTC in La Palma. This high-speed camera uses four dichroic beamsplitters and five frame-transfer CCDs to simultaneously image the $u_\mathrm{s}$, $g_\mathrm{s}$, $r_\mathrm{s}$, $i_\mathrm{s}$ and $z_\mathrm{s}$ optical bands\footnote{HiPERCAM is equipped with ''Super'' SDSS filters, high-throughput versions of the SDSS filters, hence the "s" subscript.}. The CCD detectors were used in full-frame mode with slow readout and no binning. The observations presented here were taken on 2021 June 15 with an exposure time of 12.9\,s (with only 8\,ms dead time between exposures) for 1129 frames ($\sim 4$-hour total coverage). The cadence in the $u_\mathrm{s}$ and $g_\mathrm{s}$ bands was slower by a factor of 15 in an attempt to increase the signal-to-noise ratio in these bands. The data were reduced using the HiPERCAM pipeline\footnote{\url{https://github.com/HiPERCAM/}}. We block averaged the $r_\mathrm{s}$, $i_\mathrm{s}$ and $z_\mathrm{s}$ images and obtained 60 images per band with improved signal-to-noise ratio. We then extracted the count rates of \Four\ and a comparison star\footnote{\href{http://vizier.u-strasbg.fr/viz-bin/VizieR-S?Gaia\%20DR2\%204153779729434598528}{Gaia\,DR2\,4153779729434598528}} via variable aperture photometry by direct summing of the sky subtracted flux over the aperture, i.e. with no profile weighting. The signal-to-noise ratios of the $u_\mathrm{s}$ and $g_\mathrm{s}$ images were insufficient to extract reliable fluxes. In fact, the source is not detected in either the $u_\mathrm{s}$ or $g_\mathrm{s}$ stacked images, so these bands are not discussed further in this paper. The measured seeing gradually degraded during the last third of the observation from a median value in the $z_\mathrm{s}$ band of 1.1 to 1.8 arcsec.

\begin{figure}
\centering
\includegraphics[width=0.9\columnwidth]{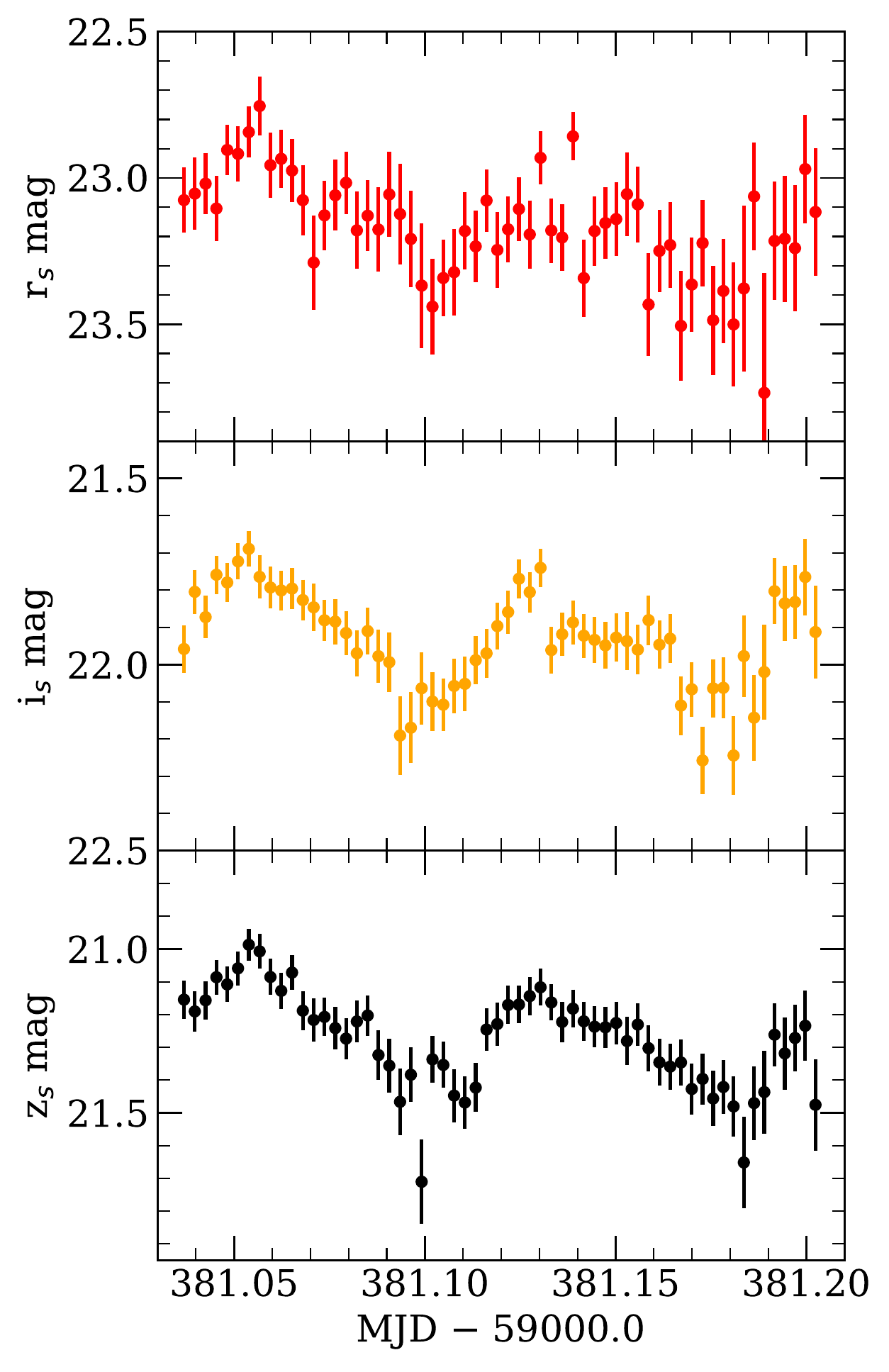}
\caption{GTC/HiPERCAM $r_\mathrm{s}$$i_\mathrm{s}$$z_\mathrm{s}$ light curves of \Four. The individual images were block averaged into 60 bins prior to flux extraction to enhance the signal-to-noise ratio.}
\label{fig:riz_lc}
\end{figure}

\section{Period analysis and results}\label{sec:Res}

The light curves of \Four\ are shown in Fig.~\ref{fig:riz_lc}. The average magnitudes of the system are $r_\mathrm{s} = 23.17 \pm 0.02$, $i_\mathrm{s} = 21.93 \pm 0.01$ and $z_\mathrm{s} = 21.28 \pm 0.01$\,mag. To compute the light curves we used the Pan-STARRS Data Release 1 magnitudes of the comparison star $r = 20.33 \pm 0.02$, $i = 18.58 \pm 0.01$ and $z = 17.42 \pm 0.01$\,mag.

\begin{figure}
\includegraphics[width=1.0\columnwidth]{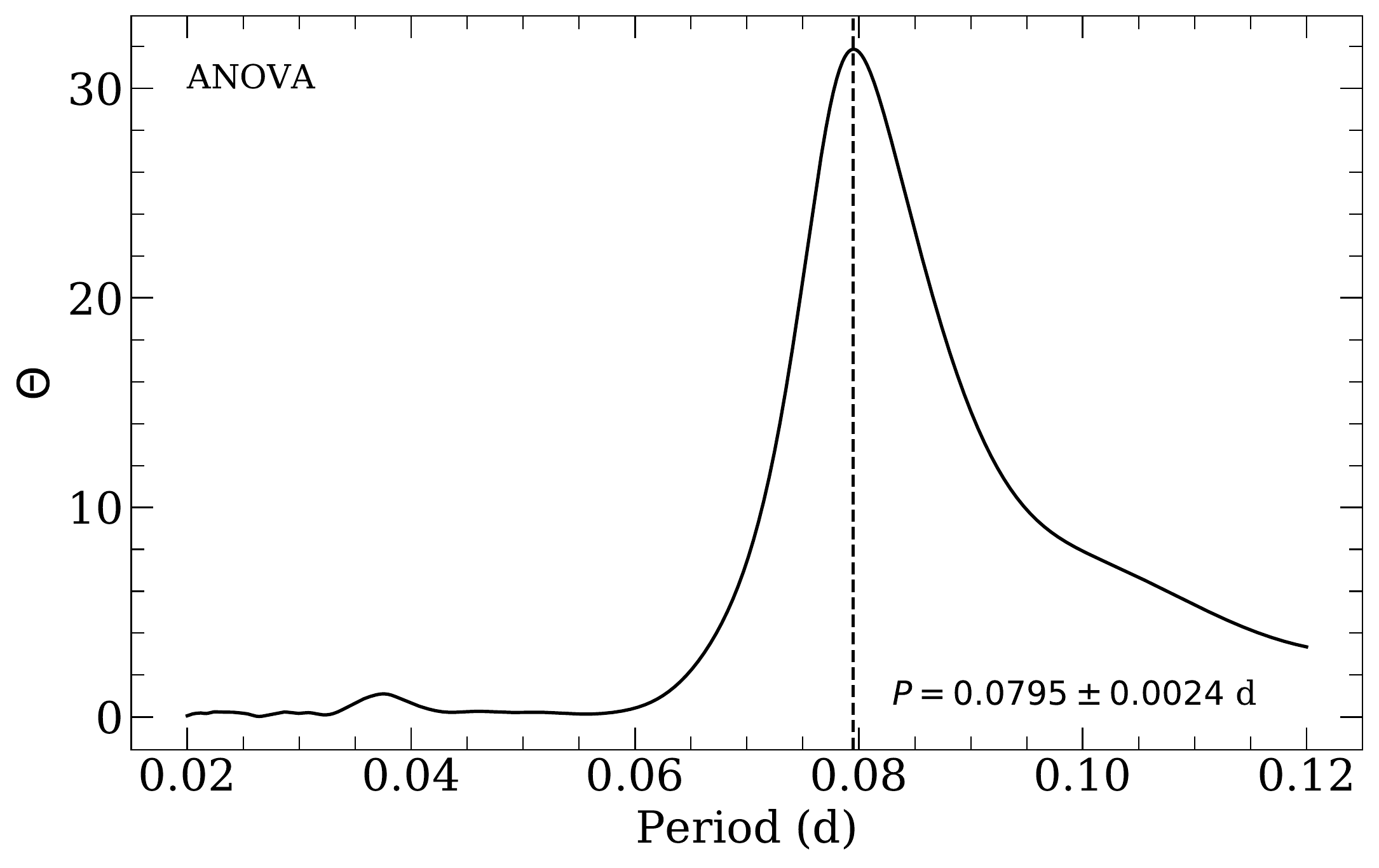}
\caption{Analysis of variance (ANOVA) periodogram of the $z_\mathrm{s}$-band light curve of \Four. A period of $114.5 \pm 3.5$~min ($0.0795 \pm 0.0024$~d) is favored.}
\label{fig:z_pgram}
\end{figure}

\begin{figure}
\centering
\includegraphics[width=0.9\columnwidth]{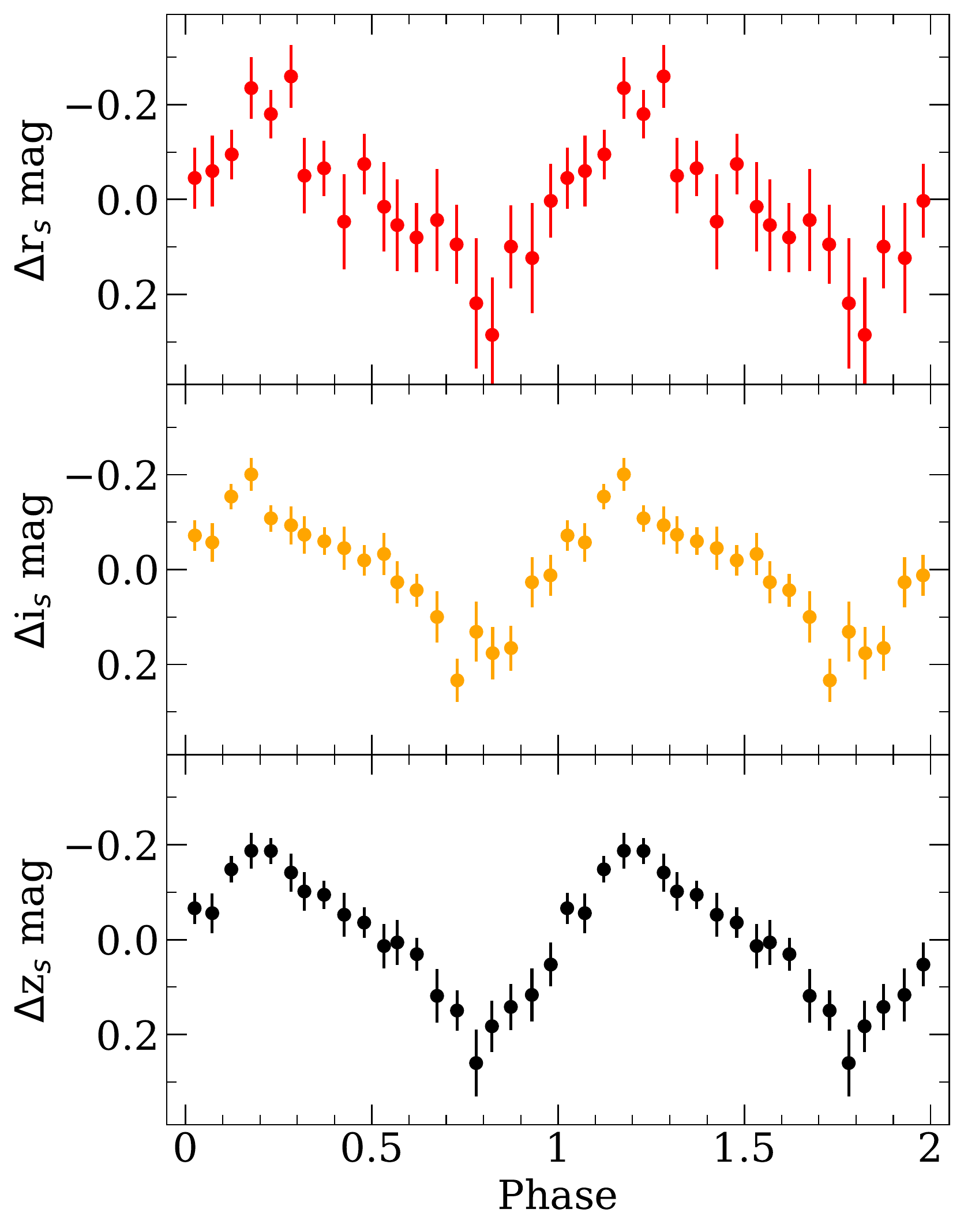}

\caption{GTC/HiPERCAM $r_\mathrm{s}$$i_\mathrm{s}$$z_\mathrm{s}$ light curves phase binned (20 bins) on the $114.5 \pm 3.5$~min period, in which the respective average magnitudes have been subtracted. The phases are computed relative to the time of the first data point. Colors are the same as in Fig.~\ref{fig:riz_lc}. The whole cycle has been plotted twice for continuity.}
\label{fig:riz_lc_folded}
\end{figure}

We analyzed the $z_\mathrm{s}$-band light curve of \Four\ using the analysis of variance (ANOVA) method \citep{schwarzenberg-czerny96-1}. This uses the ANOVA statistic, $\Theta$, to assess the goodness of fits to the data with periodic orthogonal polynomials. The periodogram is displayed in Fig.~\ref{fig:z_pgram} and favors a period of $114.5 \pm 3.5$~min ($0.0795 \pm 0.0024$~d), where the 1-$\sigma$ confidence interval was calculated using a postmortem analysis \citep{schwarzenberg-czerny91-1} and is defined as the width of the periodogram peak at the mean noise power level (MNPL; $\Theta_\mathrm{MNPL} = 2.52$ in this case) in its vicinity. We also used the ANOVA method with the $r_\mathrm{s}$ and $i_\mathrm{s}$ light curves and obtained consistent results. In Fig.~\ref{fig:riz_lc_folded} we show the average-subtracted light curves phase binned on this period. A peak-to-peak amplitude of $\approx 0.4$~mag is found for the three bands.

The observed photometric modulation can be a reflection of the orbital period and may be produced by either X-ray irradiation of the donor star or a superhump with a period a few per cent longer than the orbital period, caused by the precession of an eccentric accretion disk \citep{VanParadijs1988, Whitehurst1991}. While X-ray heating of the donor produces sinusoidal light curves   \citep[e.g.,][]{Paradijs1995}, changes in the disk size and shape and resonance between the Keplerian orbits in the disk and the orbital motion of the donor star can produce a more complex morphology in the superhump modulations \citep{Odonoghue1996, Haswell2001, Zurita2008}.

The sawtooth-like modulation that we detected (Fig.~\ref{fig:riz_lc}) supports the superhump scenario, and the phase-binned light curves presented in Fig.~\ref{fig:riz_lc_folded} are similar to the superhumps observed in cataclysmic variables (CVs; \citealt{Patterson2005}) and X-ray binaries \citep{Odonoghue1996,Zurita2002,Zurita2008}. Furthermore, the modulation is found to be color independent, as is also the case for superhumps (\citealt{Zurita2008} and references therein). Persistent CVs show permanent superhumps with typical amplitudes (peak-to-peak) of $\approx0.1$~mag \citep{Smak2010}. These are smaller than the $\approx 0.4$~mag amplitude in \Four\ (Fig. \ref{fig:riz_lc_folded}). However, superhumps with larger amplitudes ($\approx 0.25-0.6$) have been observed during CV superoutbursts \citep{Smak2010}. These are also consistent with those found in superhumps detected during LMXB outbursts ($\approx 0.1$ to 0.6~mag; \citealt{Odonoghue1996, Zurita2008,Thomas2022a}) and permanent superhumps in persistent LMXBs. For instance, a $\approx 0.6$~mag amplitude permanent superhump has been detected in the ultra-compact X-ray binary 4U~1915$-$05 \citep{Callanan1995, Chou2001, Haswell2001, Retter2002}, which has a low persistent luminosity, similar to \Four.

It is important to bear in mind, however, that our observation spans two and a half periods, and therefore we can not fully discard a non-coherent origin for the modulation, although we deem it unlikely. Rapid aperiodic variability with a suggested origin in the accretion disk is commonly observed in X-ray binaries \citep{Shahbaz2003, Zurita2003, Hynes2004b, Shahbaz2013, Casares2014}. However, this flickering activity results in erratic variations of typically very short time scales (from seconds to minutes). Longer flares with time scales of hours have also been observed, particularly in systems with long orbital periods (e.g., the 6-h flares in the 6.5-d orbital period LMXB V404~Cyg, \citealt{Casares2014}). However, such a long orbital period would require a much higher mass-accretion rate in \Four\ to sustain its persistent nature, which would not be consistent with its observed X-ray luminosity. Flaring events produced by the compact jet are also unlikely, since the jet is not expected to make a dominant contribution to the optical emission in neutron star X-ray binaries \citep[e.g.,][]{Russell2006,Migliari2010}. Further, they should be more prominent in the red bands \citep[e.g.,][]{Gandhi2016}. All considered, although we cannot fully discard that the modulation may result from the detection of two $\approx 114$~min long, color-independent and remarkably similar (in shape and amplitude) flares, a superhump origin for the observed modulation seems to be the most plausible scenario. We therefore tentatively propose an orbital period \Po\ $\approx 114$~min for \Four.\footnote{Negative superhumps, with a period slightly shorter than the \Porb, have also been detected in some systems, and are suggested to be produced by retrograde precession of a tilted disk.}

\section{Discussion}\label{sec:Disc}

\Four\ is a strong ultra-compact X-ray binary candidate. It shows several of the distinctive characteristics of the class, related to their compact geometry (small accretion disk) and nature (hydrogen-exhausted donor; see more details in \citealt{IntZand2007}; \citealt{ArmasPadilla2019b}). In particular, since its discovery 50 years ago by the Uhuru mission \citep{Forman1976}, the system has been repeatedly detected in the X-rays at $\sim$~(3--10)~$\times10^{-10}~\flux$ (depending on the X-ray band), which together with the suggested distance of $2.3-4.6$~kpc  \citep{Cocchi2000, Jonker2004,Galloway2020}, translates to a persistent X-ray luminosity of $\sim$~0.1--1~$\times 10^{36}~\lum$ \citep[e.g.][]{Warwick1981, Barret2003, Wilson2003, Muno2005b, Tarana2006}. This persistent activity at such low X-ray luminosity prompted \citet{IntZand2007} to propose an ultra-compact orbit for the source, since only small disks can be entirely ionized at such low accretion rates \citep{Lasota2001}. In the same way, \citet{Bassa2006} suggested a short orbital period based on the very dim optical counterpart ($g \simeq 25$, $r \simeq 23$\,mag; \citealt{ArmasPadilla2020}), since the X-ray (to optical) reprocessing scales with the size of the accretion disk \citep{Paradijs1994}. Yet, possibly the most compelling evidence for the ultra-compact nature of \Four\ is the absence of hydrogen features in its optical spectrum \citep{ArmasPadilla2020}, in particular \ha, which is typically the most prominent optical emission line in LMXBs with hydrogen-rich donor stars \citep{Charles2006}.

\begin{figure}$\phantom{!t}$
\centering
\begin{tabular}{c}
\includegraphics[keepaspectratio,width=\columnwidth, trim=1.0cm 0.0cm 0.0cm 1.0cm]{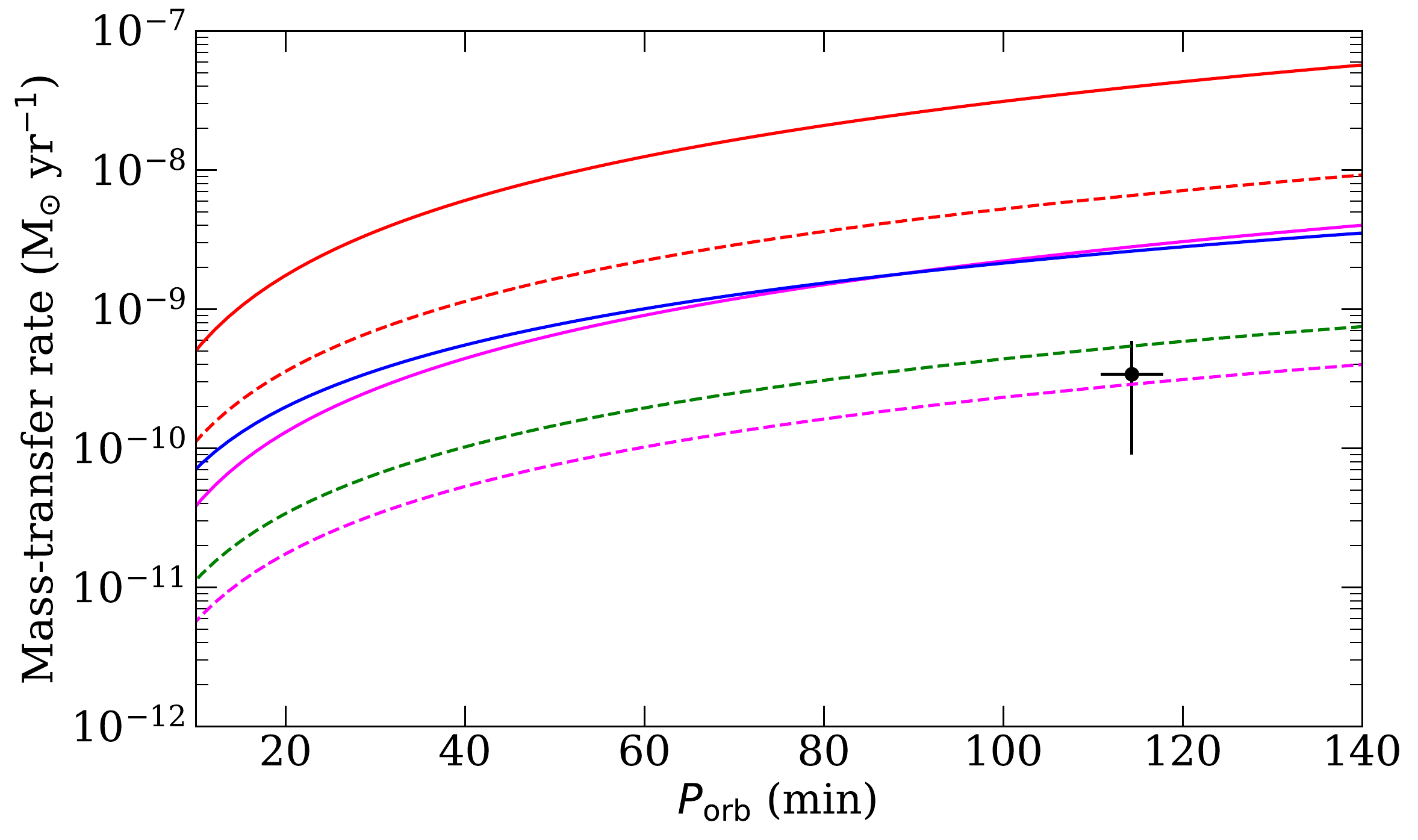}

\end{tabular}
\caption{Stability limits for non-irradiated pure-helium disks (red solid line), for irradiated pure-helium disks (red dashed line), for non-irradiated C/O disks (blue solid line), for irradiated mixed-composition disks (green dashed line), for non-irradiated solar-composition disks (magenta solid line) and for irradiated solar-composition disks (magenta dashed line) according to \citet{Menou2002} and \citet{Lasota2008}. The black dot corresponds to \Four\ assuming the tentative $\approx 114$-min orbital period and the mass transfer rate reported in this work (see Section~\ref{sec:Disc}).}
\label{fig:DIM}
\end{figure}


If we assume that \Four\ is indeed an ultra-compact system (i.e., hydrogen poor), the numerous short thermonuclear bursts displayed by the source would need to be fuelled by pure He, pointing to a He-rich donor (see \citealt{ArmasPadilla2020}). Considering evolutionary tracks for donor stars in ultra-compact binaries, the derived mass transfer rate\footnote{We derived the mass transfer rate following \citet{Coriat2012}, assuming an average 2--10 keV X-ray unabsorbed continuum flux of $3.8\times10^{-10}~\flux$ \citep{Cocchi2000}, a bolometric correction of 2.9 \citep[for which we account a 25 per cent uncertainty;][]{IntZand2007} and a distance of $2.3-4.6$~kpc \citep{Jonker2004, Galloway2020}.} of $\sim (3.4 \pm 2.5)\times10^{-10}\,\Msun$~yr$^{-1}$ translates into an orbital period of $\simeq 20$~min and $\gtrsim 35$~min for a He white dwarf and a He-star donor, respectively. However, according to disk instability models for irradiated He accretion disks, the orbital period should be $\lesssim25$~min in order to sustain a stable disk ($\lesssim40$~min in the case of a C/O disk; \citealt{Menou2002, Lasota2008}).

\subsection{An ultra-compact X-ray binary progenitor?}\label{subsec:evolved}
Our tentative \Po\ $\approx 114$~min lies outside the ultra-compact zone of \Po~$<80$~min, contrary to what was expected from the aforementioned observational properties. 
The binary orbital separation is still large enough to fit a main sequence star. In fact, a 114-min orbit would require a very late M-type main sequence star or a brown dwarf to fill its Roche lobe \citep{Faulkner1972, Cox2000, Rappaport2021}. In such cases, the system mass ratio would be $q=M_{2}/M_{1} < 0.04$ (assuming  $M_{1}=1.4\,\Msun$), which is in agreement with the low mass ratio required to produce superhump modulations \citep{Whitehurst1991}. Further,  according to evolutionary sequences for LMXBs, systems with an initial companion star mass $\approx 0.6-3\,\Msun$ and an orbital period below the bifurcation period evolve by shrinking the orbit to a minimum period that can be as low as \Porb\ $\simeq 80$~min \citep[see e.g.,][]{Rappaport1982, Podsiadlowski2002}.
Both scenarios could be valid for \Four. However, the lack of hydrogen features in its optical spectrum is difficult to reconcile with these binary solutions. Nevertheless, we note that some transient LMXBs with hydrogen-rich companions did not show \ha\ emission during some phases of the outburst  (see \citealt{Jimenez-Ibarra2019, Stoop2021} and references therein).  

A very appealing alternative for the nature of the system is that \Four\ is a progenitor of an ultra-compact X-ray binary. In this case, the donor star would be an evolved main-sequence star that started mass transfer near or just after the point of central hydrogen exhaustion, and is progressively getting closer to the compact object on its path to the ultra-compact period regime (i.e., the system is in the so-called evolved main-sequence star channel; see \citealt{Nelemans2010b}). In this scenario, traces of hydrogen can still be present in the donor star photosphere ($X_{\rm s}\sim 0.1$), and therefore in the accretion disk \citep{Nelemans2010b, Nelson2003, Podsiadlowski2002}. This scenario would reconcile all the observables of \Four.  First, this solution would be consistent with it being a persistent system, since the mass transfer rate is of the same order as the critical mass transfer rate in an accretion disk with mixed composition ($X=0.1$ and $Y=0.9$, \citealt{Lasota2008}, see Fig.~\ref{fig:DIM}). Second, such a low fraction of hydrogen would not affect the duration of thermonuclear bursts \citep{Cumming2003}, which is in agreement with the numerous short thermonuclear events displayed by the source \citep[][]{Cocchi2000, Galloway2020}. Finally, this low fraction of hydrogen would not be detectable in the optical spectrum \citep{Werner2006}, in agreement with the observations of \citet{ArmasPadilla2020}.

\subsection{An ultra-compact X-ray binary with a 114-min orbital period?}\label{subsec:ucxb}
According to evolutionary models, ultra-compact X-ray binaries can evolve towards orbital periods of $100-110$~min or longer if the donor is heated and inflated, or if the donor’s mass loss via winds is taken into account \citep{VanHaaften2012a, VanHaaften2012b}. However, the predicted mass transfer rate at these longer periods is below $\sim 10^{-12}~\mdot$, which is two orders of magnitude lower than that of \Four. We note that these computed tracks are for a helium or carbon-oxygen white dwarf companion that fully fills its Roche lobe. The evolutionary tracks for He-star donors presented in \citet{Heinke2013} provide higher mass transfer rates, which may explain the group of persistent ultra-compact systems with orbital periods longer than $\simeq 40$~min and high mass accretion rates. Still, these tracks were calculated only for orbital periods up to 60~min. Thus, the evolution beyond this value is unclear, as is the stage where the He-star core stops expanding. 

Setting aside the uncertainties that surround the above He-star evolutionary paths, this tantalizing scenario (i.e., an evolved 114-min ultra-compact system) might be still plausible from a pure stability point of view. Although the mass transfer rate of \Four\ sits below the disk instability lines for both pure He and non-irradiated C/O disks, an irradiated C/O disk would maintain stability if the critical mass transfer rate drops by a similar amount than He and Solar-abundance disks when irradiation is taken into account (i.e., by a factor of $6-10$, see Fig.~\ref{fig:DIM}, \citealt{Menou2002,Lasota2008}). As a matter of fact, evolutionary calculations show that He-star donors can be C/O rich stars with some traces of helium left \citep{Nelemans2010b}. Furthermore, we note that our calculated mass transfer rate is an upper limit, since it does not account for possible mass loss via outflows (see e.g., \citealt{Fender2016, Santisteban2019, Marino2019b}), and therefore the actual mass-transfer rate might still sit above the instability thresholds.

\section{Conclusions}\label{sec:concl}

We have presented time-resolved HiPERCAM/GTC photometry  of the ultra-compact X-ray binary candidate \Four. The $r_\mathrm{s}$$i_\mathrm{s}$$z_\mathrm{s}$ light curves show a clear sawtooth-like periodic modulation that resembles a superhump. An  ANOVA periodogram of the $z_\mathrm{s}$-band light curve favors a period of $\simeq 114$~min. We tentatively propose this as the orbital period of \Four, challenging the previously proposed ultra-compact nature of the binary.

We discuss possible scenarios for the nature of the system.  Based on its properties, namely its persistently dim luminosity, optical spectrum and short thermonuclear type I bursts, we suggest that \Four\ is an ultra-compact X-ray binary progenitor whose orbit is shrinking towards the ultra-compact regime and has an evolved main-sequence star with a low fraction of photospheric hydrogen as donor star.

Additional high-quality observations of \Four\ are desirable in order to verify the persistence of the periodic modulation and confirm the orbital period of the system.

\begin{acknowledgments}
This work is supported by the Spanish Ministry of Science under grants PID2020--120323GB--I00 and EUR2021--122010. We acknowledge support from the Consejería de Economía, Conocimiento y Empleo del Gobierno de Canarias and the European Regional Development Fund (ERDF) under grant with reference ProID2021010132 and ProID2020010104. The design and construction of HiPERCAM was funded by the European Research Council under the European Union’s Seventh Framework Programme (FP/2007--2013) under ERC--2013--ADG Grant Agreement no. 340040 (HiPERCAM). HiPERCAM operations and VSD are supported by STFC grant ST/V000853/1. Based on observations made with the Gran Telescopio Canarias (GTC), installed at the Spanish Observatorio del Roque de los Muchachos of the Instituto de Astrofísica de Canarias, on the island of La Palma. This research made use of Peranso (\url{www.peranso.com}), a light curve and period analysis software.

\end{acknowledgments}

%

\vspace{5mm}
\facilities{GTC(HiPERCAM)}


\software{ANOVA \citep{schwarzenberg-czerny96-1},  
          HiPERCAM pipeline (\url{https://github.com/HiPERCAM/}),
          Peranso (\url{www.peranso.com})
          }

\bibliography{4U1812-12_Period.bbl}{}
\bibliographystyle{aasjournal}



\end{document}

%% file: definitions.tex
\newcommand{\ha}{H$\alpha$}

\newcommand{\Po}{$P_\mathrm{orb}$}

\newcommand{\Msun}{\mathrm{M}_{\odot}}
\newcommand{\lum}{\mathrm{erg~s}^{-1}}
\newcommand{\flux}{\mathrm{erg~cm}^{-2}~\mathrm{s}^{-1}}

\newcommand{\mdot}{\mathrm{M}_{\odot}~\mathrm{yr}^{-1}}